\documentclass[twocolumn]{article}
\usepackage{graphicx}
\usepackage{amsmath}
\usepackage{hyperref}
\usepackage{amsfonts}
\usepackage{amssymb}
\advance\textheight by \topskip
\def\widetext{\par\global\columnwidth42.5pc
\global\hsize\columnwidth\global\linewidth\columnwidth
\global\displaywidth\columnwidth}
\def\footnoterule{\kern-19pt\hrule width.5in\kern18.6pt}
\begin{document}
\title{On the deformation of spontaneously twisted fluctuating ribbons}
\author{Sergey Panyukov$^{*}$ and Yitzhak Rabin\dag}
\date{{\small \textit{$^{*}$Theoretical department, Lebedev Physics Institute,
Russian Academy of Sciences, Moscow 117924, Russia; \hspace{2.5cm} and
$^{\dag}$Department of Physics, Bar--Ilan University, Ramat-Gan 52900, Israel}}}
\maketitle

\textbf{ ABSTRACT A theoretical analysis of the effect of force and torque on
spontaneously twisted, fluctuating elastic ribbons is presented. We find that
when a filament with a straight center line and a spontaneously twisted
noncircular cross section is subjected to a sufficiently strong extensional
force, its average elongation exhibits an asymmetric response to large over
and undertwist. We construct the stability diagram that describes the buckling
transition of such ribbons under the opposing action of force and torque and
show that all the predicted behaviors can be understood in terms of continuous
transformations between straight and spiral/helical states of the ribbon. The
relation between our results and experimental observations on DNA is discussed
and a new reentrant spiral to rod transition is predicted at intermediate
values of twist rigidity and applied force.}

\noindent\hrulefill

The first experiments on the stretching of double stranded DNA
(dsDNA)\cite{SFB92} were successfully described by the wormlike chain
model\cite{MS94,BMSS94} that accounts for the bending rigidity of the molecule
. While an extension of this model reproduced the observed bell-shaped curves
that characterize the response of dsDNA to torque in the weak stretching
regime\cite{BM98}, it could not capture the observed under/overtwist asymmetry
at larger applied force\cite{SABBC96,SCB98}. The feeling that linear
elasticity alone can not account for the above observations, led to the
development of a semi--microscopic theory in which the two strands of the
double helix were modeled as elastic filaments connected by rigid rods that
represent the base pairs, and an energy penalty was introduced for folding the
base pairs with respect to the central axis of the double helix\cite{Haijun}.
This theory was able to reproduce most of the experimental observations on the
deformation of dsDNA. However, the question is whether one is indeed forced to
resort to such microscopic models to account for the experimental findings. In
this paper we show that the linear theory of elasticity of spontaneously
twisted ribbons combined with statistical mechanics, reproduces the
qualitative features of the experiments on stretching and twisting of dsDNA,
and provides important insights about the physical origin of the observed phenomena.

\section{THE MODEL}

The configuration of a thin ribbon of length $l$ and asymmetric cross section
is described by a triad of unit vectors $\left\{  \mathbf{t}_{i}(s)\right\}
$, where $s$ ($0\leq s\leq l$) is the contour distance, $\mathbf{t}_{3}$ is
the tangent to the center line and $\mathbf{t}_{1}$ and $\mathbf{t}_{2}$ are
oriented along the principal axes of the cross section. The orientation of the
triad as one moves along the filament is given by the solution of the
generalized Frenet equations that describe the rotation of the triad vectors,
$\mathbf{t}_{i}^{\prime}(s)=\sum_{j,k}\varepsilon_{ijk}\omega_{j}%
(s)\mathbf{t}_{k}(s)$, where $\varepsilon_{ijk}$ is the antisymmetric tensor,
the prime denotes differentiation with respect to $s$ ($\mathbf{t}_{i}%
^{\prime}=d\mathbf{t}_{i}/ds)$, and $\left\{  \omega_{i}(s)\right\}  $ are the
curvature and torsion parameters\cite{Helix}. These equations can be rewritten
in terms of the Euler angles $\theta,\varphi$ and $\psi$:
\begin{align}
\theta^{\prime}  &  =\omega_{1}\sin\psi+\omega_{2}\cos\psi,\nonumber\\
\varphi^{\prime}\sin\theta &  =-\omega_{1}\cos\psi+\omega_{2}\sin
\psi,\label{dAng}\\
\psi^{\prime}\sin\theta &  =(\omega_{1}\cos\psi-\omega_{2}\sin\psi)\cos
\theta+\omega_{3}\sin\theta.\nonumber
\end{align}
Notice that while the angles $\theta(s)$ and $\varphi(s)$ describe the spatial
conformation of the center line, $\psi(s)$ describes the rotation of the cross
section about this center line.

The elastic energy of a deformed ribbon is given by the sum of bending and
twist contributions, $U_{el}=U_{bend}+U_{twist}$ where\cite{Helix}
\begin{equation}%
\begin{array}
[c]{c}%
U_{bend}=\dfrac{kT}{2}%
%TCIMACRO{\dint _{0}^{l}}%
%BeginExpansion
{\displaystyle\int_{0}^{l}}
%EndExpansion
ds\left(  a_{1}\omega_{1}^{2}+a_{2}\omega_{2}^{2}\right)  ,\\
U_{twist}=\dfrac{kT}{2}a_{3}%
%TCIMACRO{\dint _{0}^{l}}%
%BeginExpansion
{\displaystyle\int_{0}^{l}}
%EndExpansion
ds\left(  \omega_{3}-\omega_{30}\right)  ^{2}%
\end{array}
\label{bend}%
\end{equation}
Here $k$ is the Boltzmann constant, $T$ is the temperature, the bare
persistence lengths $a_{1}$ and $a_{2}$ represent the bending rigidities with
respect to the two principal axes of inertia of the cross section, and $a_{3}$
is associated with twist rigidity. We assume that the stress-free reference
state corresponds to a ribbon with a straight center line oriented along the
$x$--axis and a cross section that is twisted about this line at a rate
$\omega_{30}$ (this defines the twist number in $Tw_{0}=l\omega_{30}/2\pi$ and
the total angle of twist $l\omega_{30}$). The above expression for the energy
is based on linear theory of elasticity and applies to deformations whose
characteristic length scale is much larger than the diameter of the
filament\cite{Love}. Since we consider the deformation of the ribbon by forces
applied to its ends, the total energy contains an additional term $-\left(
kTf/l\right)  \int_{0}^{l}ds\sin\theta\cos\varphi$ where $f$ is the force in
units of $kT/l$. This theory is a generalization of the wormlike chain
model\cite{MS94} to the case of a ribbon with noncircular cross section and
nonvanishing spontaneous twist.

In this work we only consider small deviations of the Euler angles
$\delta\theta$, $\delta\varphi$ from their values $\theta_{0}=\pi/2$,
$\varphi_{0}=0$ in the stress-free state. Physically, this corresponds to the
strong force regime, $f\gg1$, in which the tangent vector $\mathbf{t}_{3}(s)$
undergoes only small fluctuations about the $x$ axis. No restrictions on the
magnitude of the deviation $\delta\psi$ from the spontaneous value $\psi
_{0}(s)=$ $\omega_{30}s$ are imposed. Expanding the energy to second order in
the deviations $\delta\theta$, $\delta\varphi$ and introducing the complex
variable $\Phi=\left[  \delta\theta+i\delta\varphi\right]  e^{-i\psi}$ yields
\begin{align}
\dfrac{U}{kT}  &  =-f+\int_{0}^{l}ds\left\{  \dfrac{f}{2l}\left|  \Phi\right|
^{2}+\dfrac{a_{1}+a_{2}}{4}\left|  \dot{\Phi}\right|  ^{2}+\right. \nonumber\\
&  \left.  \dfrac{a_{2}-a_{1}}{8}\left[  \dot{\Phi}^{2}+\left(  \dot{\Phi
}^{\ast}\right)  ^{2}\right]  \right\}  +\dfrac{U_{twist}}{kT}, \label{Utot}%
\end{align}
where we defined $\dot{\Phi}=\Phi^{\prime}+i\psi^{\prime}\Phi,$ and where
$\omega_{3}=\psi^{\prime}+\frac{i}{4}\left(  \Phi^{\ast}\dot{\Phi}-\Phi
\dot{\Phi}^{\ast}\right)  $.

We proceed to calculate the free energy of the deformed ribbon, for a given
value of the total angle of twist, $\psi(s)=\psi^{\prime}s,$ where we assume
that rate of change $\psi^{\prime}$ is constant (a priori, we do not exclude
the possibility that solutions with different values of this constant coexist
in the filament). The constrained partition function is evaluated by carrying
out the functional integrals over $\Phi$, $\Phi^{\ast}$ with the Boltzmann
weight $\exp(-U/kT)$. Since $U_{twist}$ contains quartic terms in these
fields, in order to calculate the integrals over $\Phi$ and $\Phi^{\ast}$ we
introduce the Hubbard--Stratonovitch transformation of the twist contribution
to the partition function
\begin{equation}%
\begin{array}
[c]{c}%
\exp\left(  -U_{twist}/kT\right)  =\int D\gamma(s)\times\\
\exp%
%TCIMACRO{\dint _{0}^{l}}%
%BeginExpansion
{\displaystyle\int_{0}^{l}}
%EndExpansion
ds\left[  \dfrac{\gamma^{2}}{2\left(  2\pi\right)  ^{2}a_{3}}-\dfrac{\gamma
}{2\pi}(\omega_{3}-\omega_{30})\right]
\end{array}
\label{hubb}%
\end{equation}
where $\gamma$ can be interpreted as a fluctuating torque. The resulting free
energy is a quadratic form in $\Phi$ and $\Phi^{\ast}$ and the Gaussian
integrals over these fields can be carried out exactly. For convenience, we
assume that the field $\Phi(s)$ obeys periodic boundary conditions,
$\Phi(L)=\Phi(0)$, and diagonalize the free energy by expanding $\Phi(s)$ in
Fourier series, $\Phi(s)=\sum_{n}\tilde{\Phi}_{n}e^{i2\pi ns/L}$. A different
choice of boundary conditions would affect our results only in the weak force
region (in this work we will only consider the range $f\gg1$). The integral
over $\gamma$ is calculated by the steepest descent method. This yields the
free energy
\begin{equation}
\dfrac{\mathcal{F}(f,\gamma)}{kT}=-f+\gamma\left(  Lk-Tw_{0}\right)
-\dfrac{\gamma^{2}}{2C}+\dfrac{1}{2}\sum_{n=-\infty}^{\infty}\ln
h_{n}(f,\gamma) \label{Free}%
\end{equation}
where we defined $Lk=\psi(l)/2\pi$ and where $h_{n}(f,\gamma)=f^{2}%
+2f[B(n^{2}+Lk^{2})-\gamma Lk]\allowbreak+(n^{2}-Lk^{2})[A^{2}(n^{2}%
\allowbreak-Lk^{2})\allowbreak+\allowbreak2B\gamma Lk-\allowbreak\gamma^{2}]$
with $A=(2\pi)^{2}\sqrt{a_{1}a_{2}}/l$, $B=(2\pi)^{2}(a_{1}+a_{2})/2l$ and
$C=(2\pi)^{2}a_{3}/l$. A relation between $Lk$ and the torque $\gamma$ is
obtained by minimizing the free energy with respect to $\gamma$:
\begin{equation}
Lk=Tw_{0}+\dfrac{\gamma}{C}+\sum_{n=-\infty}^{\infty}\dfrac{(n^{2}%
-Lk^{2})\left(  \gamma-BLk\right)  +Lkf}{h_{n}(f,\gamma)}. \label{SCR}%
\end{equation}
This relation has a simple geometrical meaning. The twist number
$Tw=(1/2\pi)\int_{0}^{l}ds\omega_{3}(s)$ can be related to the torque $\gamma$
by steepest descent evaluation of the integral in Eq. (\ref{hubb}) that yields
$Tw=Tw_{0}+\gamma/C.$ Inverting Eq. (\ref{dAng}) and assuming $|\delta
\theta|\ll1$, the twist number can be written as $Tw=(1/2\pi)\int_{0}%
^{l}ds\left[  \psi^{\prime}-\delta\theta d\left(  \delta\varphi\right)
/ds\right]  $. For $|\delta\theta|\ll1$ the writhe number can be expressed
as\cite{BM98} $Wr=(1/2\pi)\int_{0}^{l}ds\delta\theta d\left(  \delta
\varphi\right)  /ds$ and thus $Tw+Wr=\psi(l)/2\pi$.$\ $Since the sum of twist
and writhe numbers is the linking number\cite{White}, we conclude that our
definition of $Lk$ coincides with the standard definition of the linking
number. Inspection of Eq. (\ref{SCR}) shows that the first two terms on the
rhs of this equation give the twist number and that the third term is the
writhe number.

A quantity that can be readily measured in experiments is the mean elongation
of the filament (average end to end distance), $\left\langle R\right\rangle
=-(l/kT)\partial\mathcal{F/}\partial f$. The dependence of the elongation on
the total angle of twist ($2\pi Lk$) and on the extensional force is studied below.

\section{RESULTS}

\begin{figure}[ht]
\includegraphics[scale=0.4,angle=-90]{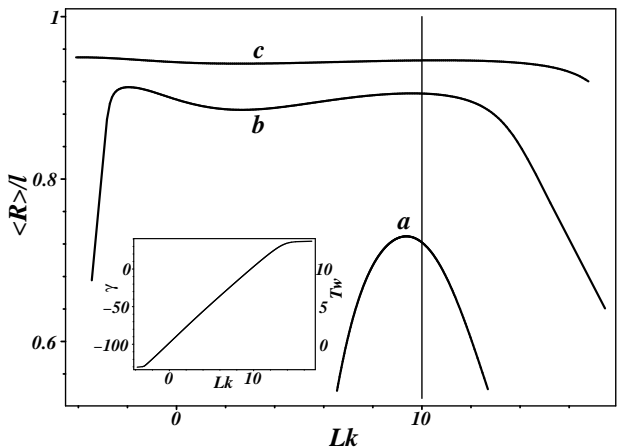}
\caption{The average normalized elongation $\left\langle
R\right\rangle /l$ is plotted vs the linking number $Lk$, for a
ribbon with $Lk_{0}=10,$ $A=10,$ $B=50$ and $C=10$
($a_{1}\simeq1,$ $a_{2}\simeq100$ and $a_{3}=10$). The
extensional force corresponding to the three curves are a) $f=$
$50,$ b) $f=$ $350$ and c) $f=$ $1000$. The dependence of the
torque (twist) on the linking number for $f=$
$350$ is shown in the insert.}%
\label{Fig1}%
\end{figure}

\textbf{Effect of torque on elongation.} In Fig.~\ref{Fig1} we plot the
average elongation $\left\langle R\right\rangle $ vs. the linking number, for
a spontaneously twisted ribbon ($Lk_{0}=Tw_{0}=10,$ $A=C=10,$ $B=50$). At
relatively low values of the force, we obtain a symmetric bell--shaped curve
(curve a), in agreement with reference \cite{BM98}. Notice that fluctuations
shift the peak of the curve to $Lk_{\max}\simeq9.4$ which is somewhat smaller
than the spontaneous value. As $f$ is increased, the dependence of the
elongation on the angle of twist becomes progressively asymmetric (curve b);
$\left\langle R\right\rangle $ decreases linearly with overtwist but is nearly
independent of undertwist throughout the range $0\lesssim Lk\lesssim$ $Lk_{0}%
$. At yet higher values of $f$, the elongation becomes nearly independent of
under or overtwist, in a broad range of linking numbers (curve c).

In order to understand how the average conformation of the filament varies
with the linking number, in the insert to Fig.~\ref{Fig1} we plot the torque
$\gamma$ and the twist number $Tw$ vs. $Lk$, for a value of $f$ that
corresponds to curve b in this figure. Both $\gamma$ and $Tw$ increase
linearly with $Lk$ in the range that corresponds to the flat portion of curve
b, and approach constant values in the range where linear decrease of
elongation with degree of over or undertwist is observed. In the range where
$Tw\ $increases linearly with $Lk$ (see insert), the writhe number remains
small and\ the application of torque to the ends of a rectilinear filament
results mainly in twist of its cross section about a straight center line. At
large values of overtwist ($\Delta Lk=Lk-Lk_{\max}>0$) for which $Tw$
approaches a constant value, further increase of the linking number leads to
the appearance of large mean writhe and the filament undergoes a transition to
a spiral or a helical configuration. As will be shown in the following, this
transition is related to buckling of elastic rods under torque\cite{Love} (in
the presence of thermal fluctuations the buckling instability is replaced by a
continuous change of shape with increasing torque). Further increase of $Lk$
increases both the amplitude (the radius) and the number of turns of the
spiral and leads to progressive shortening of the average length of the
filament, in agreement with the behavior observed in Fig.~\ref{Fig1}b.
Inspection of the insert shows that linear variation of $Tw$ with $Lk$ takes
place over much broader range of undertwist than of overtwist, indicating that
throughout the plateau region in Fig.~\ref{Fig1}b, the removal of spontaneous
twist takes place mainly by untwisting a straight filament, leaving
$\left\langle R\right\rangle $ unaffected. The shallow minimum in the plateau
region of curve b is related to the existence of a reentrant spiral/helix to
rod transition that will be discussed below.

\begin{figure}[h]
\includegraphics[scale=0.45,angle=-90]{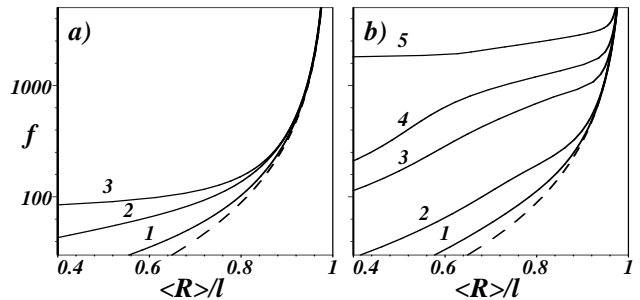}
\caption{a) Plot of $f$ vs. $\left\langle R\right\rangle /l$, for
increasing degrees of undertwist. The broken line corresponds to
a spontaneously twisted ribbon ($Lk=Lk_{0}$), and curves $1-3$ to
$\Delta Lk=-1.5$, $-3$ and $-10,$ respectively. b) Plot of $f$
vs. $\left\langle R\right\rangle /l$ for increasing degrees of
overtwist. The broken line corresponds to $Lk=Lk_{0}$, and curves
$1-5$ to $\Delta Lk=1.5,$ $3$, $7$, $10$ and $15,$ respectively.
All elastic parameters are as in Fig.
\ref{Fig1}.}%
\label{Fig2}%
\end{figure}

\textbf{Force versus elongation.} In Figs.~\ref{Fig2}a and~\ref{Fig2}b we plot
$f$ vs. $\left\langle R\right\rangle $ for different values of $Lk$
corresponding to undertwist and overtwist, respectively. For small deviations
from $Lk_{0}$, the curves are symmetric under $\Delta Lk\leftrightarrow-\Delta
Lk$. For larger deviations a plateau-like region is observed at intermediate
elongations, that is reached more rapidly for undertwist than for overtwist.
Since for each set of parameters we find a unique solution $\psi^{\prime}$ of
Eq. (\ref{SCR}), the above plateau is not associated with coexistence of two
phases. Rather, the observation of a region of nearly constant $\log f$ over a
large range of elongations means that in this region of parameters $f$ varies
slower than exponentially with $\left\langle R\right\rangle $ (e.g., as a power).

\begin{figure}[h]
\includegraphics[scale=0.4,angle=-90]{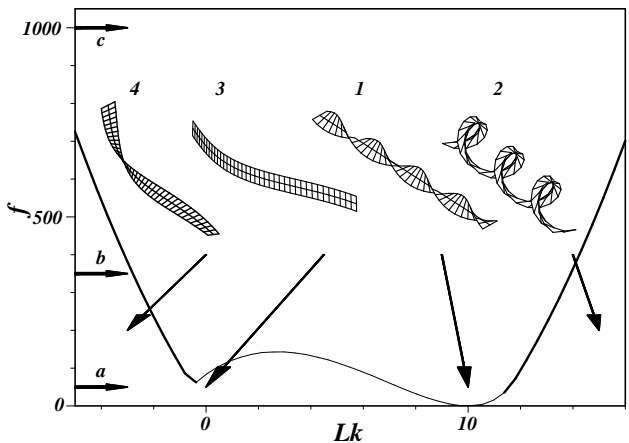}
\caption{The stability diagram in $f$ vs $Lk$ plane, for elastic
ribbon with the same parameters as in Fig. \ref{Fig1} (the
straight twisted ribbon configuration is stable against buckling 
above this line).
The configurations of the filament in
different regions of the diagram indicated by the arrows are shown in inserts 
$1-4$. The
arrows denoted by a, b and c on the lefthand side of the figure
correspond to the appropriate curves in
Fig. \ref{Fig1}.}%
\label{Fig3}%
\end{figure}

\textbf{Buckling under torque.} The simple physical picture behind the above
observations is intimately related to the well-known buckling transition under
torque and, in order to gain intuition about the different parameter ranges,
we examine the stability of the straight filament under the combined action of
tension and torque. We substitute Eq. (\ref{hubb}) into $\exp(-U/kT)$ and look
for the conditions under which the quadratic form in $\Phi$ is no longer
positive definite. In the absence of thermal fluctuations, a straight filament
becomes unstable against buckling under torque at $h_{n}(f,\gamma)=0.$ The
shape of the filament following buckling is given by a combination of Fourier
modes $\tilde{\Phi}_{n}$ and $\tilde{\Phi}_{-n}$, where $n$ is found by
minimizing $h_{n}(f,\gamma)$ with respect to $n$. In the presence of thermal
fluctuations $h_{n}(f,\gamma)>0$ everywhere and the buckling instability is
replaced by a continuous deformation of the mean conformation of the filament
from a straight line into a three-dimensional curve. In the limit of strong
force and torque (the regime in which $\left\langle R\right\rangle $ decreases
linearly with $Lk-$see curve b in Fig. \ref{Fig1}), the resulting shape is
dominated by a combination of modes with wave vectors $2\pi\left(  Lk\pm
n\right)  /l$. The case $n=0$ corresponds to a simple helix with period $l/Lk$
and the smaller axis of inertia of the cross section (direction of easy
bending) rotating in the plane normal to its symmetry axis (see Fig.
\ref{Fig3}).

In Fig. \ref{Fig3} we plot the line of buckling transitions in the $f-$ $Lk$
plane for the same elastic parameters as in the preceding figures. The thick
portions of the solid line correspond to instability at $n=0$ (rod to helix
transition) and the thin portion of the line corresponds to $n\neq0$ (rod to
spiral transition). The four inserts describe the average configuration of the
ribbon, at the locations shown by the corresponding arrows on the stability
diagram. Insert $1$ describes a ribbon with a cross section spontaneously
twisted about a straight center line (no applied torque), and corresponds to
the stable region above the minimum at $Lk=Lk_{0}$. Insert $2$ corresponds to
the unstable region to the right of this minimum (overtwist), in which
buckling transforms the straight twisted ribbon into a helix, with the smaller
moment of inertia of its cross section oriented normal to the axis of symmetry
of the helix. The origin of the broad symmetric minimum at $Lk=Lk_{0}$ is
intuitively clear--small over/undertwist with respect to the equilibrium
configuration destabilizes the filament against buckling and a larger
extensional force is needed to maintain the straight state. The fact that in
the absence of torque the instability appears to take place at $f\rightarrow0$
rather at some negative value of the stretching force corresponding to the
Euler instability under load, is a consequence of our periodic boundary
conditions; since the ends of the filament are not fixed, for $f\rightarrow0$
there is an instability against rigid rotation of the filament. The smaller
minimum at $Lk\simeq0$ appears only in the presence of large spontaneous twist
($Lk\gg1$) and bending asymmetry ($a_{1},a_{3}\ll a_{2}$). Insert $3$ shows a
typical configuration in the unstable region just below this minimum --
buckling produces an untwisted ribbon bent along its easy axis. Since the
stretching of this filament is opposed only by the smaller of the bending
rigidities, it can be easily stretched into a straight configuration. The
presence of even small deviations from $Lk=0$ creates a nonplanar
configuration shown in insert $4$, the stretching of which invokes both the
easy ($a_{1}$) and the hard ($a_{2}$) bending axes, and requires a larger
extensional force (hence the minimum).

The arrows a, b, c on the left hand side of Fig. \ref{Fig3} refer to the
corresponding curves in Fig. \ref{Fig1}. Recall that while thermal
fluctuations were neglected in the derivation of the stability diagram in Fig.
\ref{Fig3}, they were included in the calculations leading to Fig. \ref{Fig1}
(their main effect is to broaden the stability line into an extended
transition region between straight and spiral/helical states of the ribbon).
Case (a) corresponds to the weak force regime in which strong fluctuations
smear out the stability curve in the vicinity of the minimum at $Lk=Lk_{0}$,
resulting in a symmetric bell-shaped curve (see Fig. \ref{Fig1}). Cases (b)
and (c) correspond to the strong force regime. For small overtwists, all
excess linking number goes into pure twist and the filament maintains (on the
average) its straight configuration. At larger linking numbers beyond the
fluctuation--broadened stability line, the filament develops positive writhe
and undergoes a continuous rod-to-helix transition. Further increase of $Lk$
increases the radius and number of turns of the helix, resulting in a linear
decrease of $\left\langle R\right\rangle $. When the ribbon is subjected to
undertwist, the filament remains straight and ``unwinds'' over a much larger
range of undertwist compared to overtwist (because of the pronounced asymmetry
of the stability line), giving rise to the plateau regions in curves b and c,
in Fig. \ref{Fig1}. Note that there is an intermediate regime between cases
(a) and (b), in which undertwist leads to reentrant behavior. For small
degrees of undertwist, decreasing $Lk$ initially unwinds the spontaneously
twisted but straight filament and, as the stability line is crossed for the
first time, the ribbon deforms into a helix with negative writhe and
$\left\langle R\right\rangle $ decreases with increasing undertwist. As the
stability line emanating from the $Lk=0$ minimum is approached, the helix
undergoes a reentrant transition into an untwisted rod and $\left\langle
R\right\rangle $ increases (a trace of this behavior is evident in curve b of
Fig. \ref{Fig1}). Finally, at yet higher undertwists the stability line is
crossed again and a transformation into a negatively twisted spiral takes
place, accompanied by rapid decrease of $\left\langle R\right\rangle $.

\section{DISCUSSION}

In this paper we studied the response of spontaneously twisted fluctuating
elastic ribbons to externally applied torque and extensional force. The
analysis is based on a combination of linear theory of elasticity and
statistical mechanics, and the only input from an underlying microscopic level
of description is contained in the values of the elastic constants and of the
rate of spontaneous twist. The agreement between our results and those of
reference \cite{Haijun}, is not surprising since the linear theory of
elasticity is the long wavelength limit of any physically reasonable
microscopic theory of solid behavior. Even though we made no effort to adjust
our model parameters to fit dsDNA\cite{units}, our Figs. \ref{Fig1} and
\ref{Fig2} contain most of the qualitative features of the experimental
observations on twisted and stretched DNA in the intermediate range of force,
$1-70$ $%
%TCIMACRO{\unit{pN}}%
%BeginExpansion
\operatorname{pN}%
%EndExpansion
$\cite{SABBC96,SCB98}. An exception to this statement is the predicted rapid
decrease of elongation in the limit of large undertwist (Fig. \ref{Fig1}b)
that was not observed in experiment, possibly because the predicted decrease
in elongation is preempted by a microscopic structural transition of dsDNA
(alternatively, the decrease in elongation may occur at yet larger degrees of
undertwist not reached in the experiments).

All the features observed in Figs. \ref{Fig1} and \ref{Fig2} can be understood
in terms of a simple physical picture based on the stability diagram, Fig.
\ref{Fig3}, that describes the buckling instability under the opposing actions
of torque and extensional force. When torque is applied to a straight
spontaneously twisted ribbon, its response depends on the direction of the
torque relative to that of spontaneous twist, and on the magnitude of $f$
(assumed to be fixed during the process). For small degrees of overtwist, the
filament remains straight and its cross section is twisted in excess of the
spontaneous value. As the transition region around the stability line is
reached, the straight ribbon deforms into a spiral curve or into a helix,
depending on $f$. When the ribbon is subjected to undertwist it unwinds while
maintaining a straight center line over much larger range than in the case of
overtwist. Eventually, the transition region is reached and the straight
ribbon deforms into a spiral or a helix. For the range of elastic parameters
studied in this paper ($a_{1}\ll a_{3}\ll a_{2}$) the stability diagram
contains an additional minimum at $Lk\simeq0$ and, for small enough $f$,
further undertwist leads to a spiral to rod transition (for forces below the
critical value corresponding to this minimum the spiral transforms into a
periodically bent planar configuration). At yet higher degrees of undertwist
the straight filament deforms again into a helix. For larger values of $f$
(above the value at the local maximum at $Lk\simeq3$ in Fig. \ref{Fig3}), the
application of undertwist leads to the untwisting of a straight filament,
resulting in a broad plateau in $\left\langle R\right\rangle $. Such behavior
was observed both in experiments on dsDNA\cite{SABBC96,SCB98} and in our Fig.
\ref{Fig1}. Note that although the value of $f$ corresponding to curve b in
Fig. \ref{Fig1} is well above the local maximum of the stability line of Fig.
\ref{Fig3}, the nonmonotonic variation of $\left\langle R\right\rangle $ with
degree of undertwist suggests that the fluctuations of the filament (and hence
its elongation) are affected by the presence of the minimum at $Lk\simeq0$.
While there is no compelling experimental evidence for reentrant behavior in
dsDNA to date (however, a shallow minimum in $\left\langle R\right\rangle $ at
intermediate values of $f$, is clearly visible in Fig. 3 of reference
\cite{SABBC96}), it will be interesting to look for it in other systems such
as dsDNA ``dressed'' by attached proteins, RNA, etc..

We would like to comment on the possible ramifications of this work. Both the
strength and the limitation of the present approach are its generality --
while the description of the deformation of spontaneously twisted ribbons is a
hitherto unsolved problem that is interesting in its own right, one may
question whether it is a suitable model for dsDNA. Thus, one may ask whether
the large asymmetry of the bending persistence lengths ($a_{2}/a_{1}\simeq
100$) necessary to produce the curves in Figs. \ref{Fig1}--\ref{Fig3} is
physically reasonable. In order to answer this question notice that if we
model a double helix by a spontaneously twisted ribbon, the bending
coefficients of this ribbon correspond to a hypothetical untwisted state of
dsDNA in which two strands connected by base pairs form a ladder--like
structure (two multiply connected, straight parallel lines). Such a structure
is expected to have large asymmetry for bending in the plane of the ladder
($a_{2}$) or normal to it ($a_{1}$) (dsDNA\ was modelled as a ribbon polymer
made of two semi-flexible chains that could not be bent in the plane of the
ribbon, in reference \cite{Liverpool}). The effective persistence length
$a_{eff}$ of our model is given by the relation $a_{eff}^{-1}=(a_{1}%
^{-1}+a_{2}^{-1})/2$ and approaches $2a_{1}$ in the limit $a_{2}\gg a_{1}$
(thus, while $a_{1}\simeq25%
%TCIMACRO{\unit{nm}}%
%BeginExpansion
\operatorname{nm}%
%EndExpansion
$ is determined by the persistence length of dsDNA, $a_{2}$ has no direct
physical interpretation).

Another issue involves the description of deformation--induced internal
transitions in dsDNA invoked by experimenters\cite{SABLC99} in order to
explain the observed plateau in plots of $f$ vs. $\left\langle R\right\rangle
$, as well as the observed opening of the DNA bases at large undertwists (by
hybridization with complementary fragments), and the increased reactivity of
some bases at large overtwists. Even though in our continuum elastic model,
there is no direct reference to internal structure, analogies between shape
changes and structural transitions can be readily drawn. Thus, variations of
internal structure would correspond to changes of the average conformation of
the filament and it is plausible that the conformations shown (see inserts
$1-4$) in Fig. \ref{Fig3} differ in their chemical reactivity, consistent with
the observations of reference \cite{SABLC99}. Furthermore, the experimental
observation that even relatively small undertwists lead to
denaturation\cite{SABLC99}, is consistent with our result that over a large
range of undertwists, the cross section unwinds around a straight center line
and no writhe develops (one expects twist around a center line to be more
effective than writhe of the center line in disrupting the interaction between
the base pairs). A deeper difference between our discussion of force and
torque induced shape changes and that of transitions between different
internal states of dsDNA in reference \cite{SABLC99} is that while we find
only continuous transformations of average conformation, discontinuous first
order internal transitions (from B to P and S forms) and phase coexistence are
invoked in the latter work.

Since our model assumes small deviations of the Euler angles from their
stress-free values, it can not describe large amplitude defects (with
$\delta\theta$, $\delta\varphi$ $\gtrsim\pi$) such as plectonemes. However,
our analysis shows that for $f\gg1$, the formation of plectonemes is preempted
by the appearance of homogeneous spiral and helical structures (recall that
the classical torsional buckling instability corresponds to a transition to a
helical state with arbitrarily small amplitude\cite{Love}), even though
plectonemes may appear at yet larger applied torques. Finally, the fundamental
limitation of our theory is that since it is based on the linear theory of
elasticity of thin inextensible rods, it can not describe the observed
overstretching of B-DNA by $1.7$ times its native length\cite{cluzel,SCB96},
and one must resort to microscopic\cite{Haijun} or
thermodynamic\cite{Bloomfield} approaches.

\medskip

{\small We would like to thank David Bensimon and M. Elbaum for helpful
comments on the manuscript. YR acknowledges support by a grant from the Israel
Science Foundation.}


\begin{thebibliography}{99}
\bibitem {units}In making the comparison one should use the force $F$ and the
degree of supercoiling $\Delta Lk/Lk_{0},$ that do not depend of the length of
the filament.
\end{thebibliography}
\end{document}